\documentclass[epj,final]{svjour}% referee final
\usepackage{graphics}
\begin{document}
\title{Solution to the non-perturbative renormalization of gauge theory}
\author{Hans-Christian Pauli}
%
%  Comments: LaTeX2e, 9 pages, 3 figures, 0 tables, 24 references. 
%
\institute{Max-Planck-Institut f\"ur Kernphysik, D-69029 Heidelberg,
           \email{pauli@mpi-hd.mpg.de}}
\date{17 December 2003}% 
\abstract{%
    The long standing problem of a non-perturbative renormalization 
    of a gauge field theoretical Hamiltonian is addressed and
    explicitly carried out within an (effective) light-cone 
    Hamiltonian approach to QCD. 
    The procedure is in line with the conventional ideas: 
    The Hamiltonian is first regulated by suitable cut-off functions,
    and subsequently renormalized by suitable counter terms to make
    it cut-off independent. Emphasized is the considerable freedom
    in the cut-off function which eventually can modify the Coulomb 
    potential of two charges at sufficiently small distances. 
    The approach provides new physical insight into the nature of 
    gauge theory and the potential energy of QCD and QED at short distance.
} 
\maketitle
%
%\tableofcontents
%
\section{Introduction} 
\label{sec:1}
One of the most interesting subjects of particle and nuclear physics is
to understand the internal structure of hadrons.  
A powerful description of the hadronic structure is obtained through 
the light-cone wave functions. 
They encode all possible quark and gluon
momenta, and helicity and flavor correlations in the hadron and are
constructed from the QCD light-cone Hamiltonian \cite{BroPauPin98}:
\begin{eqnarray*} 
   H_\mathrm{LC} = P^+P^- - P^2_{\!\perp}
\;,\end{eqnarray*} 
where $P^{\pm} = P^0 \pm P^z$. 
The wave function 
$\Psi_b$ of a hadron $b$ with mass $M_b$ 
satisfies the relation \cite{BroPauPin98}:
\begin{eqnarray*} 
   H_\mathrm{LC}\vert\Psi_b\rangle = M_b^2 \vert\Psi_b\rangle
\;.\end{eqnarray*} 
The light-cone wave functions are expanded in terms of a complete basis of 
Fock states having increasing complexity. In this way the hadron 
presents itself as an ensemble of coherent states containing various
numbers of quark and gluon quanta. For example, the positive 
pion has the Fock expansion:
\begin{eqnarray*}
   \vert\Psi_{\pi^+}\rangle 
   &=& \sum _n \langle n\vert\pi^+\rangle \vert n\rangle  
   = \Psi^{(\Lambda)}_{u\bar d/\pi}(x_i,\vec{k}_{\!\perp i}) 
     \vert u\bar d\rangle 
\\ &+&   
   \Psi^{(\Lambda)}_{u\bar dg/\pi}(x_i,\vec{k}_{\!\perp i})
   \vert u\bar dg\rangle  + \dots 
\;,\end{eqnarray*}
representing an expansion of the exact QCD eigenstate at scale $\Lambda$ 
in terms of non-interacting quarks and gluons.
The $i_n$ particles in a Fock state ($n$) have 
longitudinal light-cone momentum fractions $x_i$
and relative transverse momenta $\vec{k}_{\!\perp i}$, with
\begin{eqnarray*}
   x_i = \frac {k_i^+}{P^+} = \frac {k_i^0+k_i^z}{P^0+P^z}\;, \quad
   \sum_{i=1}^{i_n} x_i =1\;, \quad
   \sum_{i=1}^{i_n} \vec{k}_{\!\perp i} = \vec 0_{\!\perp} 
\;.\end{eqnarray*}
The form of $\Psi_{n/H}(x_i,\vec{k}_{\!\perp i})$ is invariant under
longitudinal and transverse boosts; i.e., the light-cone wave functions 
expressed in the relative coordinates $x_i$ and $k_{\!\perp i}$ are 
independent of the total momentum ($P^+$, $\vec P_{\!\perp}$) of the hadron. 
The first term in the expansion is referred to as the valence Fock state,
as it relates to the hadronic description in the constituent quark model.
The higher terms are related to the sea components of the hadronic 
structure.
It has been shown that the remining terms of the light-cone wave function 
can be determined once the valence Fock state is known \cite{Mue94,Pau99b}.
Explicit expressions are given in \cite{Pau99b}.

When starting in 1984 with 
Discretized Light-Cone Quantization (DLCQ) \cite{PauBro85a} 
and with a
revival of Dirac's Hamiltonian front form dynamics \cite{dir49}, 
all challenges of a gauge field Hamiltonian theory 
were essentially open questions, particularly 
\begin{itemize}
\item[$\bullet$]
   the non-perturbative bound state problem,
%  Lorenz and gauge invariance, 
\item[$\bullet$]
   the many-body aspects of gauge field theory,   
\item[$\bullet$]
   regularization,  
\item[$\bullet$]
   renormalization,  
\item[$\bullet$]
   confinement, 
\item[$\bullet$]
   chirality,
\item[$\bullet$]
   vacuum structure and condensates,
\end{itemize}
just to name a few. 
The step from the gauge field QCD Lagrangian
down to a non-relativistic Schr\"odinger equation 
was completely mysterious.
Now we know better \cite{BroPauPin98}. 

We have understood, for example, that 
the chiral phase transition, in which the quarks are supposed
to get their mass, is not the major challenge.
The challenge is to understand what happens \emph{after} 
the phase transition, at zero temperature.

The challenge is to understand the spectra of physical hadrons
and to get the corresponding eigenfunctions, the light-cone wave functions. 

We have learned how to partition the problem and how 
to shape our thinking in four major steps:
\begin{eqnarray} 
   &\mathcal{L}_\mathrm{QCD}&  
\nonumber\\ 
   &\Downarrow&
\nonumber\\
   H_\mathrm{LC} \vert \Psi \rangle &=& M^2 \vert \Psi \rangle
\label{eq:lch}\\ 
   &\Downarrow&
\nonumber\\
   H_\mathrm{eLC}\vert \Psi _{q\bar q}\rangle 
   &=& M^2 \vert \Psi _{q\bar q}\rangle
\label{eq:elch}\\ 
   &\Downarrow&
\nonumber\\
   H_\mathrm{eff}\;\vert \Psi _{q\bar q}\rangle 
   &=& E\;\vert \Psi _{q\bar q}\rangle 
\label{eq:momsp}\\  
   &\Downarrow&
\nonumber\\
   \left(\frac{\vec p^{\,2}}{2m_r}+V(\vec x)\right)\psi(\vec x) 
   &=& E\;\psi(\vec x)
\;.\label{eq:consp}\end{eqnarray}
The step from the Lagrangean $\mathcal{L}_\mathrm{QCD}$ 
to the light-cone Hamiltonian $H_\mathrm{LC}$ had been given
in \cite{BroPauPin98}; all its matrix elements are known and tabulated.
Its reduction to an effective light-cone Hamiltonian $H_\mathrm{LC}$ 
has been given in \cite{Pau99b,Pau98} and will be sketched once again in
Sec.~\ref{sec:4}. Some time will be spent in Sec.~\ref{sec:5} 
to discuss the single particle Hamiltonian $H_\mathrm{eff}$ 
in momentum and in configuration space.
%
%%%%%%%%%%%%%%%%%%%%%%%%%%%%%%%%%%%%%%%%%%%%%%%%%%%%%%%%%%%%%% beg figure
\begin{figure*}\sidecaption
   \resizebox{0.24\textwidth}{!}{\includegraphics{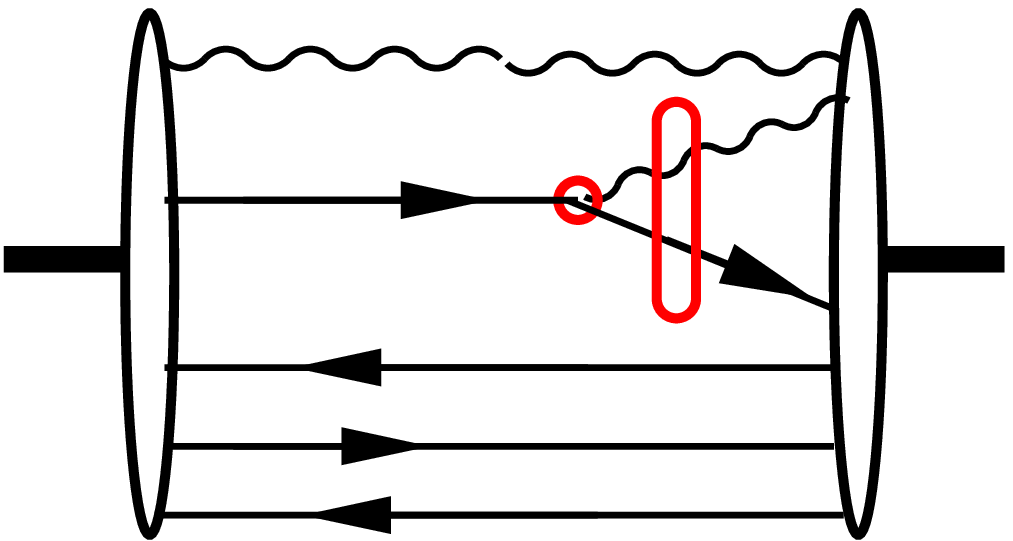}
}  \resizebox{0.38\textwidth}{!}{\includegraphics{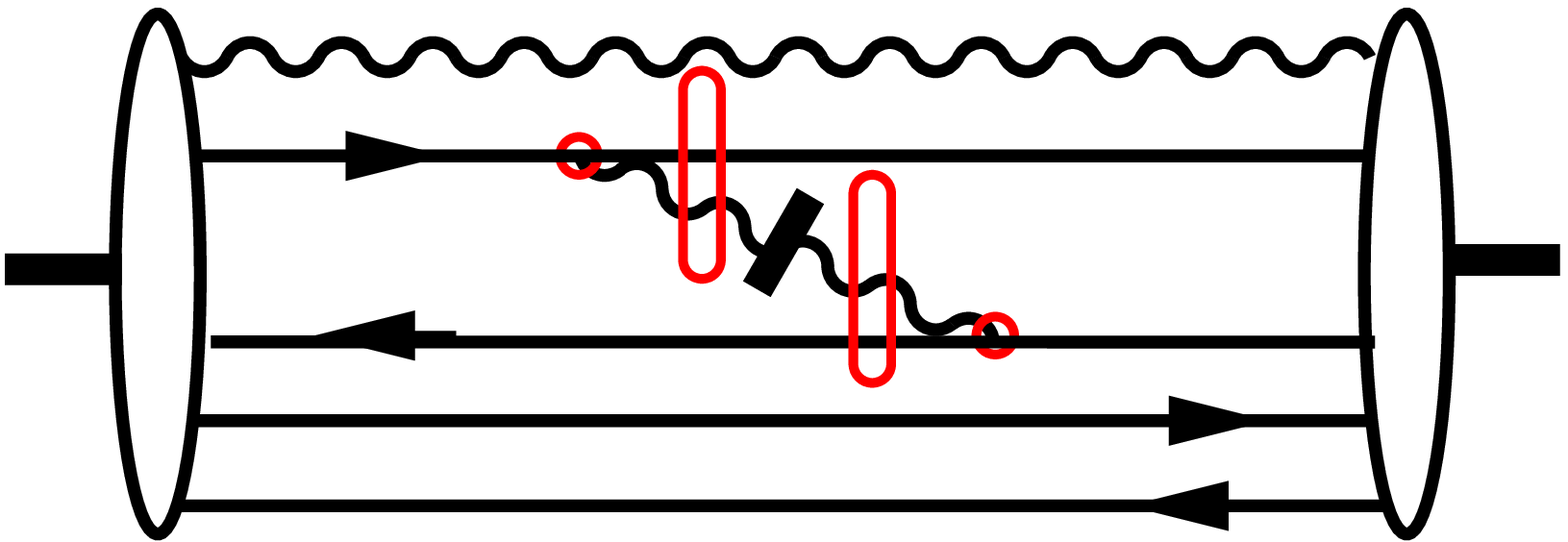}
}\caption{ 
   Regularization of the interaction by vertex regularization. 
   In a matrix element, as illustrated on the left for a vertex,
   a quark changes its four-momentum from $k_1$ to $k_2$, 
   \textit{i.e.} $Q^2= -(k_1-k_2)^2$. 
   The vertex interaction is regulated by multiplying with a 
   form factor $F(Q)$, as indicated by the circle.~--
   Instantaneous interactions are treated correspondingly,
   as illustration on the right for a seagull.
}\label{fig:reg}\end{figure*}
%%%%%%%%%%%%%%%%%%%%%%%%%%%%%%%%%%%%%%%%%%%%%%%%%%%%%%%%%%%%%% end figure
%

But first we must deal with the key issue of any gauge theory.
We must say something about how to treat the logarithmic divergencies
of the theory.
We shall do it by conventional regularization and renormalization
in Secs.~\ref{sec:6} and \ref{sec:7}.
In the perturbative context of scattering theory 
the problem of regularization and renormalization was solved. 
In the non-perturbative context of a Hamiltonian
it has not been solved. 
It is addressed in the next two sections. 

Preliminary and condensed versions of this were given
in conference proceedings \cite{Pau03a}.
Several typos and misprints are removed, also including some mistakes.
\section{Regularization of the total Hamiltonian}
\label{sec:2}
Canonical field theory with the conventional QCD Lagrangian 
allows to derive the components of the total canonical 
four-momentum $P^\mu$.
Its front form version \cite{BroPauPin98} rests on two assumptions, 
the light cone gauge $A^+=0$ \cite{LepBro80} and the
suppression of all zero modes \cite{BroPauPin98,Kal95}.
The front form vacuum is then trivial.

I find it helpful to discuss the problem in terms of 
DLCQ \cite{BroPauPin98,PauBro85a}.
In the back of my mind I visualize an explicit finite dimensional 
matrix representation of the Light-Cone Hamiltonian 
as it occurs for finite harmonic resolution. 
Such one is schematically displayed in Fig.~2 of \cite{BroPauPin98}. 
All of its matrix elements are
finite for any finite $x$ and $k_{\!\perp}$. 

The problem arises for ever increasing harmonic resolution,
on the way to the continuum limit: 
The numerical eigenvalues 
are numerically unstable and diverge logarithmically \cite{KraPauWoe92,TriPau00},
in contrast to the calculations in 1+1 dimension \cite{PauBro85a};
see also actual DLCQ calculations in 3+1 by Hiller \cite{Hil00}.

The reason is inherent to Dirac's   
relativistic vertex interaction 
$\langle k_1,h_1\vert V \vert k_2,h_2;k_3,h_3\rangle$, 
in which some particle `1' is scattered into two particles `2' and `3'
with their respective four-momenta $k$ and helicities $h$,
see Fig.~\ref{fig:reg}.
The matrix element for bremsstrahlung, for example,
is proportional to $k _{\!\perp}$,
$  \langle k_1,\uparrow\vert V \vert k_2,\uparrow;
   k_1,\uparrow\rangle 
   \propto \vert\vec k _{\!\perp}\vert$, 
see Table~9 in \cite{BroPauPin98},
when the quark maintains its helicity 
while irradiating a gluon with four-momentum 
$k_3^\mu=(xP^+,\vec k _{\!\perp},k_3^-)$. 
Singularities typically arise when over the squares of such matrix elements 
is integrated with repect to $\vec k _{\!\perp}$ as in the  
integrations of perturbation theory.

As usual, regularization is not unique, and can be done in many ways.
But not all regularization techniques of the past are applicable
in a Hamiltonian approach. 
Dimensional regularization, for example, is not applicable
in a matrix approach which is stuck with the precisely 3+1 dimensions of
the physical world.
Perturbative regularization \cite{BroRosSua73} is not
applicable in the non-perturbative context.
The Fock space regularization of Lepage and Brodsky \cite{LepBro80}, 
see also \cite{BroPauPin98}, has blocked renormalization for many years.
It was impossible to find suitable counter terms.
In recent work, the invariant mass squared regularization has been
abandoned in favor of Pauli-Villars regularization
\cite{Hil00,BroHilMcC}. 
But thus far it is unclear how the approach \cite{BroHilMcC}
is applicable to the spectra of physical mesons.
After applauding the light-cone approach \cite{wil89},
Wilson and collaborators \cite{WilWalHar94} have attempted to
base their considerations almost entirely 
on a renormalization group analysis,
but no specific technology has emerged thus far.

In recent years I have come to favor another regularization,
\emph{vertex regularization}, 
which allows for renormalization and which is sufficiently simple 
on the technical level to be carried out explicitly.

The singularities are avoided \textit{a priori}
by vertex regularization, by multiplying each (typically off-diagonal) 
matrix element with a regulating \emph{form factor} $F$:
\begin{eqnarray}
   &&
   \langle k_1,h_1\vert V \vert k_2,h_2;k_3,h_3\rangle \;\Longrightarrow
\nonumber\\ 
   && 
   \langle k_1,h_1\vert V \vert k_2,h_2;k_3,h_3\rangle \; F(Q) 
\;.\label{eq:2}\end{eqnarray}
It took me several years to realize that it is the Feynman 
four-momentum transfer, $Q^2 = -(k_1-k_2)^2$, across a vertex, 
which governs an effective interaction. 
The minimal requirement for such a universal form factor is then
\begin{equation} 
    F(Q)=F(Q;\Lambda)=\left\{
    \begin{array}{ll}
      1\;,&\mbox{ for } Q^2 \rightarrow0     \;,\\  
      0\;,&\mbox{ for } Q^2 \rightarrow\infty\;.
    \end{array}\right.
\label{eq:3}\end{equation}
It is satisfied by the step function, 
$F(Q)=\Theta(Q^2-\Lambda^2)$.
The limit $\Lambda\rightarrow0$ suppresses the interaction
all together, the limit $\Lambda\rightarrow\infty$
restores the interaction and its problems. 
Any finite value of $\Lambda^2$ restricts $Q^2$ to be finite 
and eliminates the singularities.
But the sharp cut-off generates problems in another corner of the theory  
and $F(Q)$ must be an analytic function of $Q$, as to be seen below. 

Vertex regularization takes thus care 
of the ultraviolet divergences.
The (light-cone) infrared singularities are taken care of as usual
by a kinematical gluon mass. 

Emphasized should be that \emph{vertex regularization} 
does not chop off individual Fock space constituents,
nor does it chop off Fock space components with large momenta,
as in Fock-space regularization \cite{BroPauPin98,LepBro80}.
Possible problems with gauge invariance are thus less severe
and unlikely.
Spoken in terms of matrices, the dimension of the
Hamiltonian matrix can be as large as it wants to be.
The limit $\Lambda\rightarrow0$ is represented by a diagonal matrix.

\section{Renormalization of the total Hamiltonian}
\label{sec:3}
The non-perturbative renormalization of the Hamiltonian 
was stuck for many years by the fact that the coupling constant $g$ 
multiplies the regulator function $F(Q)$ in Eq.(\ref{eq:2}).
It has always been clear that one may add \emph{non local counter terms} 
\cite{WilWalHar94}, but is was not clear how they could be constructed.
An explicit construction is given below.

What follows is the general but abstract procedure 
of modern renormalization theory \cite{Schwinger00}.

Suppose to have solved the problem for a fixed value of 
the 7 `bare' parameters in the Lagrangian,
the coupling constant $g=g_0$ and the 6 flavor quark masses $m_f=(m_f)_0$,
and for a fixed value of the exterior cut-off scale $\Lambda=\Lambda_0$.
Suppose further that these 7+1 parameters are determined such,
that the calculated mass squared eigenvalues $M^2_i$ 
agree with the corresponding experimental numbers.
Next, suppose to change the cut-off  by a small amount $\delta\Lambda$.
Every calculated eigenvalue will then change by $\delta M^2_i$.
Renormalization theory is then the attempt to reformulate the
Hamiltonian, such, that all changes $\delta M^2_i$ vanish identically.

The fundamental renormalization group equation is therefore:
\begin{equation}
   \left. d M^2_i \right\vert_{0}=
   \left. d M^2_i \right\vert_{g=g_0,m_f=m_{f_0},\Lambda=\Lambda_0}=0 
\;,\label{eq:4}\end{equation}
\emph{for all eigenstates} $i$. 
Equivalently one requires that 
\emph{the Hamiltonian is stationary} with respect to small $\delta \Lambda$:
\begin{eqnarray}
   \left.\phantom{M^2_i}\delta  H_\mathrm{LC}\right\vert_{0} = 0
\;.\label{eq:5}\end{eqnarray}
Hence forward reference to ($g_0,m_{f_0},\Lambda_0$), 
to the `renormalization point', will be suppressed.

Next, I specify a non local counter term, as follows.
To the original Hamiltonian one adds a counter term Hamiltonian, 
whose interaction has exactly the same structure 
except that the regulator function $F(Q)$ is replaced
by an other function $\overline F(Q)$, that is by
\begin{eqnarray}
    F(Q,\Lambda) \Longrightarrow \overline F(Q,\Lambda) =
    F(Q,\Lambda) + C(Q,\Lambda)
\;.\label{eq:7}\end{eqnarray}
Here, the regulating function $F(Q)=F(Q,\Lambda)$ is
supposed to be a given function. 
The counter term $C(Q)=C(Q,\Lambda)$ is unknown, 
subject to the constraint that it vanishes at the 
renormalization point,
\begin{eqnarray}
    \left.\phantom{\frac{d\overline R}{d\Lambda}}
    C(Q,\Lambda)\right|_{\Lambda=\Lambda_0}=0
\;.\label{eq:8}\end{eqnarray} 
The goal is then to find a $C(Q,\Lambda)$ such that Eq.(\ref{eq:5}) 
is satisfied identically.

According to Eq.(\ref{eq:2}) the light-cone Hamiltonian depends on 
$\Lambda$ only through the function $\overline F (Q;\Lambda)$.
Its variation with respect to $\Lambda$ is therefore
\begin{eqnarray*}
   \delta H_\mathrm{LC} = 
   \delta \overline F \frac{\delta H_\mathrm{LC}}{\delta\overline F} 
\;.\end{eqnarray*}
Eq.(\ref{eq:4}) as the fundamental equation of renormalization theory 
can therefore be replaced by
\begin{eqnarray}
   \delta \overline F = \delta \Lambda
   \frac{\partial\overline F }{\partial\Lambda}= 0 
\;,\label{eq:6a}\end{eqnarray}
since the variational derivative of the Hamiltionian with respect 
to the regulator cannot vanish identically. 

The fundamental equation (\ref{eq:6a}) defines then a differential equation
for $C(\Lambda)=C(Q;\Lambda)$:
\begin{equation}
    \frac{dC(Q;\Lambda)}{d\Lambda} = - \frac{dF(Q;\Lambda)}{d\Lambda} 
\;,\end{equation} 
which, in its  integral form, includes the initial condition 
\begin{equation}
    C(Q,\Lambda) = - \int\limits_{\Lambda_0}^{\Lambda}
    \!\!ds\ \frac{dF(Q,s)}{ds} = 
    F(Q,\Lambda_0) - F(Q,\Lambda) 
\;.\label{eq:9}\end{equation}
The renormalized regulator function 
is \emph{manifestly independent of $\Lambda$}: 
\begin{eqnarray}
    \overline F(Q,\Lambda) &=& F(Q,\Lambda) + \hspace{8ex}C(Q,\Lambda) 
\nonumber\\  &=& 
    F(Q,\Lambda) + \left(F(Q,\Lambda_0) - F(Q,\Lambda)\right)
\nonumber\\  &=& 
    F(Q,\Lambda_0)
\;.\label{eq:10}\end{eqnarray}
The counter term improved Hamiltonian is renormalization group invariant.
By construction, the value of $\Lambda_0$ must be determined by experiment.

One should emphasize an important point:
In deriving Eq.(\ref{eq:10}), it was assumed implicitly 
that the regulator function has well defined derivatives 
with respect to $\Lambda$.
The theta function of the sharp cut-off, however, 
is a distribution with only ill defined derivatives;
therefore, it must be removed from the class of admitted regulator functions.

Another point is equally important:
The step function has only a single parameter, $\Lambda$.
Renormalization theory admits \emph{any function} $F(Q,\Lambda)=f(Q/\Lambda)$
as long as $f(x)$ has well defined derivatives with respect to
$x=Q/\Lambda$.

For a long time, the main result of this section looked to me
like sheer witchcraft: One selects a particular value of 
$\Lambda=\Lambda_0$, and then it turns out that the Hamiltonian is 
stationary with respect to small changes in $\Lambda$,
for any value of $\Lambda_0$!  

How can one believe this strange result? Is it only formal?~---
The abstract procedure remains somewhat academic since 
a reliable numerical solution of the bound state equation,
Eq.(\ref{eq:lch}), 
would exhaust a large fraction of the worlds computer power,
not to speak about the many variations which should be performed. 
To carry out such a programme would probably be impossible.
Progress has come only from recent work on a particular model 
which did allow to formulate a paradigmatic example \cite{FrewerFrePau02} 
in modern renormalization theory, see below.

The result is in sharp contrast to perturbative renomalization,
where the renormalized quantity is \emph{independent of the
regularization scale}. Here,
in the non-perturbative renormalization of a Hamiltonian, 
the regularization scale $\Lambda_0$ is fixed from the outset and
becomes a member of the canonical parameters, 
like mass and coupling constant.
It may not be driven to the extreme, to infinity or to zero, 
it must be kept at the finite value fixed \textit{ab initio}.

\section{The effective (light-cone) Hamiltonian}
\label{sec:4}
In a field theory, one is confronted with a many-body problem
of the worst kind: Not even the particle number is conserved.
In order to formulate effective Hamiltonians more systematically,
a novel many-body technique had to be developed,
the \emph{method of iterated resolvents} \cite{Pau99b,Pau98},
whose details are not important here.

Important is that 
the method of iterated resolvents does not chop off 
Fock states or Fock state classes,
and that it \emph{does not violate on purpose} 
symmetries like gauge and Lorentz invariance. 

Important is as well that 
the \emph{effective light-cone Hamiltonian} $H_\mathrm{eLC}$
has the same eigenvalue as 
the \emph{full light-cone Hamiltonian} $H_\mathrm{LC}$ 
and that it generates the bound state wave function of valence quarks
by an one-body integral equation
in  ($x,\vec k_{\!\perp}$) \cite{Pau98}:
\begin{eqnarray} 
\lefteqn{\hspace{-2em}
    M^2\psi_{h_1h_2}(x,\vec k_{\!\perp}) = 
    \left[ 
    \frac{\overline m^2_{1} + \vec k_{\!\perp}^{\,2}}{x} +
    \frac{\overline m^2_{2} + \vec k_{\!\perp}^{\,2}}{1-x}  
    \right]
    \psi_{h_1h_2}(x,\vec k_{\!\perp})  
}\nonumber\\ 
    &-& {1\over 3\pi^2}
    \sum _{ h_q^\prime,h_{\bar q}^\prime}
    \!\int\!
    \frac{dx^\prime d^2 \vec k_{\!\perp}^\prime}
    {\sqrt{ x(1-x) x'(1-x')}}
    \;\psi_{h_1'h_2'}(x',\vec k_{\!\perp}')
\nonumber\\  
    &\times&
    F(Q_{q}) F(Q_{\bar q})
    \left(\frac{\overline\alpha(Q_{q})}{2Q_{q}^2} +
    \frac{\overline\alpha(Q_{\bar q})}{2Q_{\bar q}^2}\right)
\nonumber\\  
    &\times&
    \left[\overline u(k_1,h_1)\gamma^\mu u(k_1',h_2')\right]
    \left[\overline v(k_2',h_2')\gamma_\mu v(k_2,h_2)\right] 
\;.\label{eq:14}\end {eqnarray}
One has arrived at Eq.(\ref{eq:elch}):
$ M^2 \vert \Psi _{q\bar q}\rangle = 
  H_\mathrm{eLC} \vert \Psi _{q\bar q}\rangle$.
Here, $M ^2$ is the eigenvalue of the invariant-mass squared. 
The associated eigenfunction $\psi_{h_1h_2}(x,\vec k_{\!\perp})$ 
is the probability amplitude 
$\langle x,\vec k_{\!\perp},h_{1};1-x,-\vec k_{\!\perp},h_{2}
\vert\Psi_{q\bar q}\rangle$ 
for finding the quark with momentum fraction $x$, 
transversal momentum $\vec k_{\!\perp}$ and helicity $h_{1}$,
and correspondingly the anti-quark.
Expressions for 
the (effective) quark masses $\overline m _1$  and $\overline m _2$ 
and the (effective) coupling function $\overline\alpha(Q)$ are given 
in \cite{Pau98}.
$Q_q$ and $Q_{\bar q}$ are the Feynman momentum transfers 
of quark and anti-quark, respectively,
and $u(k_1,h_1)$ and $v(k_2,h_2)$ are their 
Dirac spinors in the Lepage-Brodsky convention \cite{LepBro80}, 
given explicitly in \cite{BroPauPin98}.
They are arranged in the Lorenz scalar spinor matrix 
\begin{eqnarray}
   \langle h_1,h_2\vert S\vert h_1',h_2'\rangle &=&
   \left[\overline u(k_1,h_1)  \gamma^\mu u(k_1',h_1')\right] 
\nonumber\\ &\times&
   \left[\overline v(k_2',h_2')\gamma_\mu v(k_2,h_2)\right]
\,\end{eqnarray}
which is a rather complicated (matrix) function of its
six arguments $x,x',\vec k_{\!\perp},\vec k_{\!\perp}'$,
as tabulated in \cite{Pau00c}.   
Finally, the form factors $F(Q)$ restrict the range of 
integration and regulate the interaction.
Note that the equation is fully relativistic and covariant.
   
It should be emphasized that Eq.(\ref{eq:14}) is valid only for 
quark and anti-quark having different flavors \cite{Pau99b,Pau98}.
The additional annihilation term for identical flavors is omitted,
but presently being investigated \cite{Kra04}.
It should also be emphasized that the same structure was obtained
by a completely different method, with 
Wegner's Hamiltonian flow equations \cite{Wegner00}.
In \cite{Wegner00} it is also shown why the concept 
of a `mean momentum transfer',
$Q^2=\frac12\left(Q_{q}^2+Q_{\bar q}^2\right)$ 
is a meaningful simplification. It allows to replace
Eq.(\ref{eq:14}) by 
\begin{eqnarray} 
\lefteqn{\hspace{-2em}
    M^2\psi_{h_1h_2}(x,\vec k_{\!\perp}) = 
    \left[ 
    \frac{\overline m^2_{1} + \vec k_{\!\perp}^{\,2}}{x} +
    \frac{\overline m^2_{2} + \vec k_{\!\perp}^{\,2}}{1-x}  
    \right]
    \psi_{h_1h_2}(x,\vec k_{\!\perp})  
}\nonumber\\ 
    &-& {1\over 3\pi^2}
    \sum _{ h_q^\prime,h_{\bar q}^\prime}
    \!\int\!\displaystyle 
    \frac{dx^\prime d^2 \vec k_{\!\perp}^\prime
    \;\psi_{h_1'h_2'}(x',\vec k_{\!\perp}')}
    {\sqrt{ x(1-x) x'(1-x')}}
    \frac{\overline\alpha(Q)}{Q^2} R(Q) 
\nonumber\\  
    &\times&
    \left[\overline u(k_1,h_1)\gamma^\mu u(k_1',h_2')\right]
    \left[\overline v(k_2',h_2')\gamma_\mu v(k_2,h_2)\right] 
\;.\label{eq:15}\end {eqnarray}
The form factors $F(Q)$ have made their way into the regulator function
\begin{eqnarray}
   R(Q)=F^2(Q)
\;.\end{eqnarray}
The same bilinear combination occurs also in the explicit expressions
for $\overline m$ and $\overline\alpha(Q)$ \cite{Pau98}, such that one
can replace the $F(Q)$'s by $R(Q)$ as an independent variable.

It is a light cone pecularity that the (light-cone) Hamiltonian $H_\mathrm{LC}$
is additive in kinetic energy terms and the interaction \cite{BroPauPin98}.
The same holds for the effective Hamiltonian $H_\mathrm{eff}$.
Since it is allowed to subtract a c-number from $H_\mathrm{eLC}$
and to divide the result by another c-number,
one can define an effective Hamiltonian $H_\mathrm{eff}$,
implicitly by
\begin{eqnarray}
   H_\mathrm{eLC}\equiv \left(\overline m_1+\overline m_2\right)^2 +
   2\left(\overline m_1+\overline m_2\right) H_\mathrm{eff}
\;,\end{eqnarray}
and explicitly by 
\begin{eqnarray}
   H_\mathrm{eff} = \frac{
   H_\mathrm{eLC}-\left(\overline m_1+\overline m_2\right)^2}{
   2\left(\overline m_1+\overline m_2\right)} 
\;.\end{eqnarray}
Such a manipulation does not change the eigenfunctions.

The eigenvalues then have the dimension of an energy
\begin{eqnarray*}
   H_\mathrm{eff}\vert \Psi _{q\bar q}\rangle = E\vert \Psi _{q\bar q}\rangle 
\;,\end{eqnarray*}
arriving this way at Eq.(\ref{eq:momsp}). 
As will be seen below, $H_\mathrm{eff}$ has much in common
with a conventional non-relativistic Hamiltonian in momentum representation,
but the approach remains fully relativistic and covariant.

Note that mass and energy in the front form, on the light cone, 
are related by
\begin{equation}
   \hspace{13ex} M^2 = 
   \left(\overline m_1+\overline m_2\right)^2 + 
   2\left(\overline m_1+\overline m_2\right) E
\;,\label{eq:22}\end{equation}
and \textbf{not by} 
\begin{eqnarray}
   \hspace{10ex} M^2 &=& 
   \left(\overline m_1+\overline m_2 + E\right)^2
\nonumber\\ &=&
   \left(\overline m_1+\overline m_2\right)^2 + 
   2\left(\overline m_1+\overline m_2\right) E+E^2
\nonumber\\ &\neq&
   \left(\overline m_1+\overline m_2\right)^2 + 
   2\left(\overline m_1+\overline m_2\right) E 
\;,\end{eqnarray}
as usual in the instant form which quantizes the system at equal usual time. 
Only if the energy is negligible, 
\textit{i.e.} only if $\left(E/(\overline m_1+\overline m_2)\right)^2\ll 1$,
the two relations coincide.
\subsection{The effective Hamiltonian in different clothes}
\label{sec:4.1}
A rather drastic technical simplification is achieved by
a transformation of the integration variable.
One can substitute the integration variable $x$
by the integration variable $k_z$, which, 
for all practical purposes,  
can be interpreted \cite{BroPauPin98} 
as the $z$-component of a 3-momentum vector
$\vec p=(k_z,\vec k_{\!\perp})$.
For equal masses $\overline m_1=\overline m_2=m$, 
the transformation is, together with its inverse,
\begin{eqnarray}
   x(k_z) &=& \frac{1}{2} \left[1+\frac{k_z}
     {\sqrt{m^2 + \vec k_{\!\perp}^{\;2}+ k_z^2}}\right]
\;.\label{eq:23}\\
   k_z^2(x) &=& (m^2+\vec k_{\!\perp}^{\,2})
   \ \frac{\left(x-\frac{1}{2}\right)^2}{x(1-x)} 
\;.\end{eqnarray}
Inserting these substitutions into Eq.(\ref{eq:15})
and defining the reduced wave function $\varphi_{h_1h_2}$ by \cite{Pau00c}
\begin{eqnarray}
   \psi_{h_1h_2}(x,\vec k _{\!\perp}) &=&  
   \frac{\sqrt{A(k_z,\vec k _{\!\perp})}}{\sqrt{x(1-x)}}
   \varphi_{h_1h_2}(k_z,\vec k _{\!\perp})    
\,,\label{eq:25}\\
   A(\vec p)&=&\sqrt{1+\frac{\vec p^{\,2}}{m^2}} 
\;,\end{eqnarray}
leads to an integral equation in 3-momentum space,
in which all reference to light-cone variables has disappeared:
\begin{eqnarray}
\lefteqn{ 
   E\varphi_{h_1h_2}(\vec p) =
   \frac{\vec p ^2}{2m_r}\varphi_{h_1h_2}(\vec p) 
}\label{eq:27}\\ &-& 
   \frac1{2\pi^2}
   \sum _{ h_q^\prime,h_{\bar q}^\prime}
   \int\!d^3p'\;\frac{\varphi_{h'_1h'_2}(\vec p')}{\sqrt{A(p)A(p')}}
   \frac{\overline\alpha_c(Q)}{Q^2} \frac{R(Q)}{4m^2} 
\nonumber\\  
   &\times&
   \left[\overline u(k_1,h_1)\gamma^\mu u(k_1',h_2')\right]
   \left[\overline v(k_2',h_2')\gamma_\mu v(k_2,h_2)\right] 
\;,\nonumber\end {eqnarray}
with $\overline\alpha_c(Q)=\frac43\overline\alpha(Q)$.
The reduced mass for $\overline m_1=\overline m_2=m$ is $m_r=m/2$.
Its Fourier transform gives Eq.(\ref{eq:consp}).

\section{Two over-simplified models}
\label{sec:5}
Krautg\"artner \textit{et al} \cite{KraPauWoe92}
and Trittmann \textit{et al} \cite{TriPau00} have shown 
how to solve an equation like Eq.(\ref{eq:15}) numerically  
with high precision. 
But since the numerical effort is considerable, 
it is reasonable to work first with an over simplified model, 
as specified next.

\textbf{The Singlet-Triplet model}.  
Quarks are at relative rest
when $\vec k _{\!\perp}= 0$ and $ x = \overline x$,
with $ \overline x \equiv \overline m_1/(\overline m_1+\overline m_2)$.
An inspection of Eq.(33) in \cite{Pau00c} reveals that   
for very small deviations from the equilibrium values, 
the spinor matrix $\langle h_1,h_2\vert S\vert h_1',h_2'\rangle$
is proportional to the unit matrix, 
\begin{eqnarray}
   \langle h_1,h_2\vert S\vert h_1'h_2'\rangle  
   &\simeq & 4 \overline m_1 \overline m_2 
   \;\delta_{h_1,h_1'}
   \;\delta_{h_2,h_2'}
\;.\end{eqnarray}
For very large deviations, particularly for
$\vec k_{\!\perp}^{\prime\,2} \gg \vec k_{\!\perp} ^{\,2}$,
holds  
\begin{eqnarray}
   Q ^2 \simeq\vec k_{\!\perp}^{\prime\,2}
   \;,\quad\mbox{and}\quad
   \langle\uparrow\downarrow\vert S\vert\uparrow\downarrow\rangle 
   \simeq 2\vec k_{\!\perp}^{\prime\,2}
\;.\label{eq:17}\end{eqnarray}
The \emph{Singlet-Triplet (ST) model} combines these aspects:
\begin{eqnarray} 
    \langle h_1,h_2\vert S\vert h_1',h_2'\rangle 
    &=&
    \delta_{h_1,h_1'}\;\delta_{h_2,h_2'}\;
    \langle h_1,h_2\vert S\vert h_1,h_2\rangle
\;,\label{eq:18}\\
    \frac{\langle h_1,h_2\vert S\vert h_1,h_2\rangle}{Q^2}
    &=&
    \left\{
    \begin{array}{ll}
    \frac{4\overline m_1\overline m_2}{Q^2}+2, 
      &\mbox{ for $h_1 = - h_2 $,}\\
    \frac{4\overline m_1\overline m_2}{Q^2},\phantom{+2} 
      &\mbox{ for $h_1 = \phantom{-} h_2 $.}\\
    \end{array}\right.
\label{eq:19}\end{eqnarray} 
For anti parallel helicities $h_1 = - h_2 $ (singlet)
the model interpolates between two extremes:
For small momentum transfer $Q$, the `2' in Eq.(\ref{eq:17}) is unimportant and 
the Coulomb aspects of the first term prevail.
For large $Q$, the Coulomb aspects are
unimportant and the hyperfine interaction is dominant.
The `2' carries the singlet-triplet mass difference.

For parallel helicities $h_1 = h_2 $ (triplets) the model 
reduces to the Coulomb kernel.
The model over emphasizes many aspects but its
simplicity has proven useful for quick and analytical calculations. 
Most importantly, the model allows to drop the helicity summations, 
simplifying the problem enormously on the technical level.

The model can not be justified in the sense of an approximation,
but it emphasizes the point that the `2', or any other constant 
in the kernel of an integral equation, leads to singularities
and thus to numerically undefined equations.
Replacing the function $\overline \alpha(Q)$ by the strong coupling constant
$\alpha_s=g^2/4\pi$ completes the model assumptions.

For the singlet, the ST-model translates Eq.(\ref{eq:27}) identically into
\begin{eqnarray}
\lefteqn{ 
   E\varphi(\vec p) =
   \frac{\vec p ^2}{2m_r}\varphi(\vec p) 
}\label{eq:32}\\ 
   &-& \frac{\alpha_c}{2\pi^2}\int\!\frac{d^3p'}{\sqrt{A(p)A(p')}}
   \left(\frac{4m^2}{Q^2} + 2 \right)
   \frac{R(Q)}{4m^2}\;\varphi(\vec p')
\;,\nonumber\end{eqnarray}
with $\alpha_c=\frac43\alpha_s$.
The equation for the triplets is obtained by dropping the `2'.
The first term in this equation, $\vec p ^2/2m_r$, coincides with 
the kinetic energy in a conventional non-relativistic Hamiltonian.
This is remarkable in view of the fact that no approximation 
to this extent has been made. 
The fully relativistic and covariant light-cone approach
has no relativistic corrections in the kinetic energy!

Since the first term in Eq.(\ref{eq:32}) is a kinetic energy,
the second must be a potential energy --- in momentum representation.
In principle, it could be Fourier transformed
with $\mathrm{e}^{-i \vec p \vec r}$ to configuration space
with the variable $\vec r$. 
But due to the factor $A(p)A(p')$ in the kernel,
the resulting potential energy would be non-local.
A Fourier transformation with $\vec Q$, \textit{i.e.} with
$\mathrm{e}^{-i \vec Q \vec r}$, would not make much sense to me.

To avoid this non-locality, I introduce the `Fourier simplification'
\begin{eqnarray}
   A(p)\equiv 1
\;.\end{eqnarray} 
With $A(p)= 1$, the mean four momentum transfer $Q^2$ reduces 
to the three momentum transfer $q^2=(\vec p-\vec p')^2$. 
In consequence, the kernel of Eq.(\ref{eq:32}),
\begin{eqnarray}
    U(\vec q) &=&  -\frac{\alpha}{2\pi^2} 
   \left(\frac{4m^2}{q^2}+2\right)
   \frac{R(q)}{4m^2}
\;,\label{eq:29}\end{eqnarray}
depends only on $\vec q=\vec p-\vec p'$.
Its Fourier transform is a local function,  
\begin{eqnarray}
   V(\vec r) &=& \int\!d^3 q\;\mathrm{e}^{-i\vec q \vec r}\;U(\vec q)
\;,\label{eq:30}\end{eqnarray}
playing the role of a conventional \emph{potential energy} 
in the Fourier transform of Eq.(\ref{eq:32}), \textit{i.e.} in
\begin{eqnarray}
   E\;\psi(\vec r) &=& 
   \left[\frac{\vec p ^2}{2m_r}+V(\vec r)\right]\psi(\vec r) 
\;.\label{eq:31}\end{eqnarray}
Here is the Schr\"odinger equation from Eq.(\ref{eq:consp}) in full glory!
Despite its conventional structure it is a front form equation, 
designed to calculate the light-cone wave function 
$\psi(\vec r)\rightarrow\varphi(\vec p)
\rightarrow\psi_{q\bar q}(x,\vec k_{\!\perp})$.

\textbf{The $\uparrow\downarrow$-model}.  
In an early stage of the present work, I thought that the Coulomb term 
in Eq.(\ref{eq:29}) needs no regularization. In consequence
I replaced the kernel by
\begin{eqnarray}
    U(\vec q) &=&  -\frac{\alpha}{2\pi^2} 
   \left(\frac{4m^2}{q^2}+2R(q)\right)
   \frac{1}{4m^2}
\;,\end{eqnarray}
referred to as the $\uparrow\downarrow$-model. 

In view of the present insight, see \cite{Pau03b}, 
this misses the point completely and is completely false.
Nevertheless, the model has played an important role in developing 
the present renormalization theory.
\section{Paradigms for renormalization theory}
\label{sec:6}
%
%
%%%%%%%%%%%%%%%%%%%%%%%%%%%%%%%%%%%%%%%%%%%%%%%%%%%%%%%%%%%%%% beg figure
\begin{figure*}\sidecaption
   \resizebox{0.30\textwidth}{!}{
   \includegraphics{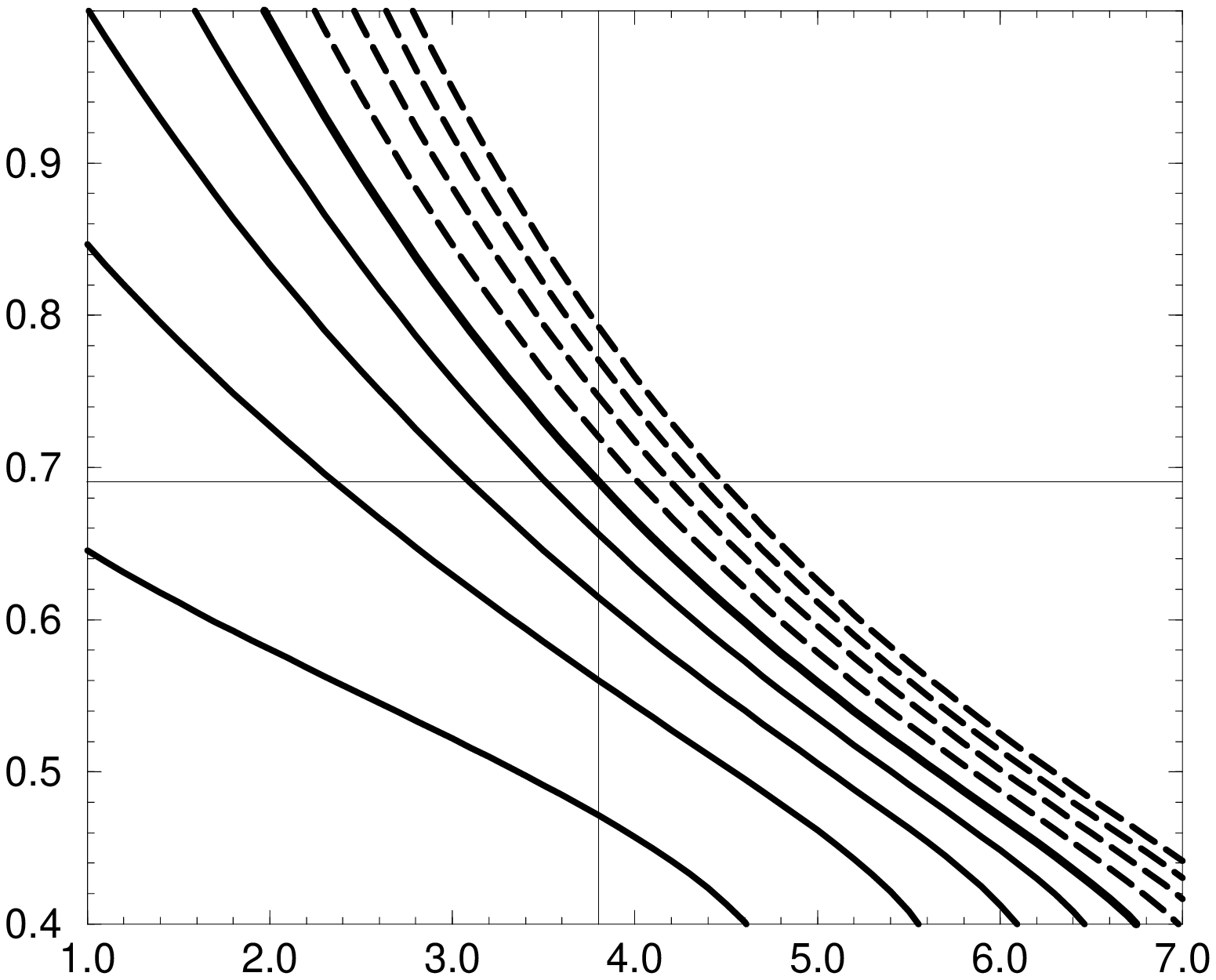}
}  \resizebox{0.30\textwidth}{!}{
   \includegraphics{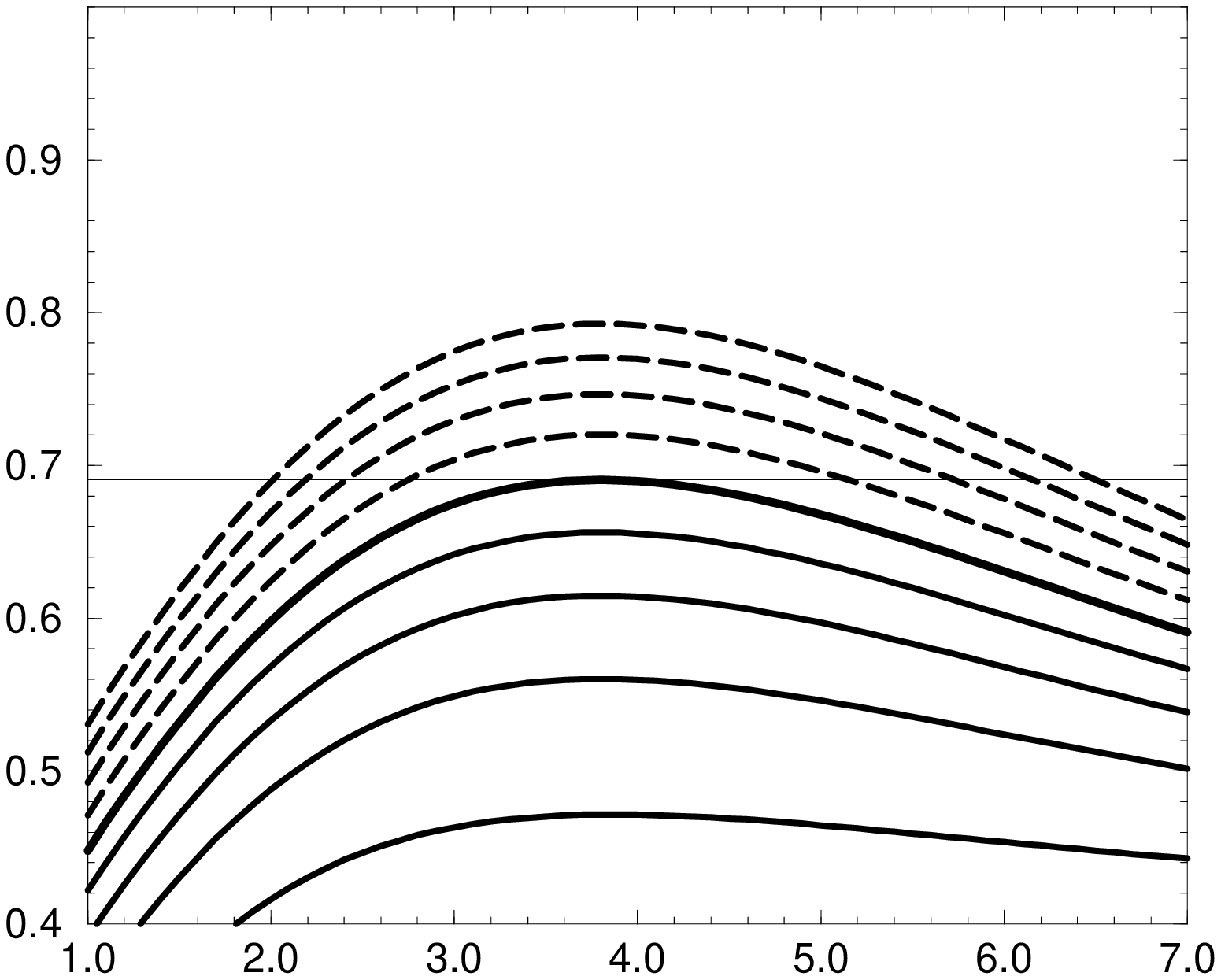}
}\caption{\label{fig:2}%
   Contour plots of $M_0^2(\alpha,\Lambda)$ versus coupling constant
   $\alpha$ and cut-off scale $\Lambda$. 
   The contours $M_0^2(\alpha,\Lambda) = n\Delta^2 + M_{\pi}^2$ 
   are denumerated with $n=4,3,\cdots ,-3,-4$ (from bottom to top).
   The thick contour $n=0$ describes the pion with 
   $M_0^2=M_{\pi}^2$.
   All masses are in units of 350~MeV.~--- 
   The horizontal lines refer to $\alpha_0=0.69$, 
   the vertical ones to $\Lambda_0=3.8$, the parameters
   which give the pion mass.~--- 
   See discussion in the text.
   Figure taken from \protect{\cite{FrewerFrePau02}}.
}\end{figure*}
%%%%%%%%%%%%%%%%%%%%%%%%%%%%%%%%%%%%%%%%%%%%%%%%%%%%%%%%%%%%%% end figure
%

In this section, I apply the $\uparrow\downarrow$-model to the
the general procedure as outlined in Sect.~\ref{sec:3}. 
It is simple enough to go to every single aspect \textit{in concreto}.
In particular, I can solve the eigenvalue problem,
$H_\mathrm{eLC}\vert\varphi_i\rangle=M_i^2\vert\varphi_i\rangle$. 

The lowest eigenvalue $M_0^2$ agrees with the pion mass at the given
values of 
\begin{eqnarray}
\begin{array}{rcl}
   \alpha=\alpha_0 &=& 0.69
\;,\\
   m = m_0 &=& 1.16 \Delta
\;,\\
   \Lambda = \Lambda_0 &=& 3.80 \Delta
\;,\end{array}
\end{eqnarray}
with the mass unit $\Delta=350\mbox{ MeV}$.
The triple will be called the renormalization point.

When changing the parameters $\alpha$ and $\Lambda$
(at a fixed value of $m = m_0$), all eigenvalues $M_i^2$ will
change and be functions of $(\alpha,\Lambda)$,
among them the mass-squared eigenvalue for the ground state, 
$M_0^2(\alpha,\Lambda)$. This function is displayed 
as a contour plot in Fig.~\ref{fig:2}(left).
Note that this plot is not schematic but the result
of an actual calculation. Calculated with 100 meshpoints
in each direction, the figure represents some 10$^4$ 
solutions of the eigenvalue problem. It goes without saying
that such can be done only for a sufficiently simple model,
the $\uparrow\downarrow$-model, 
with a sufficiently simple soft regulator 
\begin{eqnarray}
   R(Q;\Lambda)=\overline R_0(Q;\Lambda)&=& 
   \frac{\Lambda^2}{\Lambda^2+Q^2}
\;,\label{eq:39}\end{eqnarray}
and with sufficiently fast numerical procedures.

According to the general outline in Sec.~\ref{sec:3}, 
one must make sure that the invariant mass squared \emph{spectrum}  
stays invariant, $\delta M_i^2(\alpha,\Lambda)=0$,  
for infinitesimal variations $\delta\Lambda$.
I do that by construction, by introducing a \emph{new regulator} 
\begin{eqnarray}
   \overline R(Q,\Lambda) = R(Q,\Lambda) - 
   \left(\Lambda^2-\Lambda_0^2\right)
   \frac{\partial R(Q,\Lambda)}{\partial \Lambda^2}
\;,\end{eqnarray} 
with the soft regulator $R(Q,\Lambda)$ given by Eq.(\ref{eq:39}). 

Both $\overline R$ and $R$ have the same dimension. 
Both are in the class of admitted functions, see Eq.(\ref{eq:3}). 
Both are equal at the renormalization point $\Lambda=\Lambda_0$: 
\begin{eqnarray}
   \overline R(Q,\Lambda_0) &=& R(Q,\Lambda_0) 
\;.\end{eqnarray}
The derivative of $\overline R$ vanishes at the renormalization 
point by construction, since
\begin{eqnarray}
   \frac{\partial \overline R(Q,\Lambda)}{\partial \Lambda^2} &=&
   2Q^2\frac{(\Lambda^2-\Lambda_0^2)}{(\Lambda^2+Q^2)^3}
\;.\end{eqnarray} 
This derivative is certainly not the zero function for all $\Lambda$,
and renormalization can be valid at most locally, 
in the vicinity of the renormalization point.

Based on the Hellmann-Feynman theorem, 
one expects a vanishing derivative of the eigenvalue 
with respect to $\Lambda$ near the renormalization point.
The actual numerical results in Fig.~\ref{fig:2}(right) illustrate this
very convincingly: All contours have zero derivatives 
at $\Lambda = \Lambda_0$.  
The Hamiltonian is locally renormalized:
$\delta M_i^2(\alpha,\Lambda)\sim 0$.
\subsection{Renormalization by a sequence of regulators}
With $\overline R_0(Q,\Lambda)$ given by Eq.(\ref{eq:39}), 
the above can be put into a systematic form. 
I can define a whole series of regulator functions 
$\overline R_n(Q,\Lambda)$ by
\begin{eqnarray}
\begin{array}{rcl}
   \overline R_1(Q,\Lambda) &=& \overline R_0(Q,\Lambda) - 
   Q^2\frac{(\Lambda^2-\Lambda_0^2)}{(\Lambda^2+Q^2)^2}
\;,\\
   \overline R_2(Q,\Lambda) &=& \overline R_1(Q,\Lambda) - 
   Q^2\frac{(\Lambda^2-\Lambda_0^2)^2}{(\Lambda^2+Q^2)^3}
\;,\\ &\dots& 
\\
   \overline R_n(Q,\Lambda) &=& \overline R_{n-1}(Q,\Lambda) - 
   Q^2\frac{(\Lambda^2-\Lambda_0^2)^n}{(\Lambda^2+Q^2)^{n+1}}
\;.\end{array} 
\end{eqnarray} 
Each member is a genuine regulator satifying the minimal
requirements of Eq,(\ref{eq:3}). All are different from 
the original regulator in Eq.(\ref{eq:39}), 
and none can be preferred over the other.
Their derivatives given by
\begin{eqnarray}
\begin{array}{rcl}
   \frac{d\overline R_1(Q,\Lambda)}{d\Lambda^2} &=& 
   2Q^2\frac{(\Lambda^2-\Lambda_0^2)}{(\Lambda^2+Q^2)^3}
\;,\nonumber\\
   \frac{d\overline R_2(Q,\Lambda)}{d\Lambda^2} &=& 
   3Q^2\frac{(\Lambda^2-\Lambda_0^2)^2}{(\Lambda^2+Q^2)^4}
\;,\nonumber\\ &\dots& 
\nonumber\\
   \frac{d\overline R_n(Q,\Lambda)}{d\Lambda^2} &=& 
   (n+1)Q^2\frac{(\Lambda^2-\Lambda_0^2)^n}{(\Lambda^2+Q^2)^{n+2}}
\;,\end{array} 
\end{eqnarray} 
they all are stationary at the renormalization point.
Carrying on the procedure to higher and higher orders,
the contours corresponding to 
Fig.~\ref{fig:2}(right) become broader and broader. 
In the limit of large order the contours become flat:
One has realized the fundamental renormalization group
equation: $d M_i^2(\alpha,\Lambda)=0$.

One can show this even analytically by considering the limit
$n\to\infty$. With
\begin{eqnarray}
   x=\frac{\Lambda^2-\Lambda_0^2}{\Lambda^2+Q^2}
\;,\end{eqnarray} 
the regulators $\overline R_n(Q,\Lambda)$ behave in the limit 
$n\to\infty$ like
\begin{eqnarray}
   \overline R_n(Q,\Lambda) &=& \overline R_0(Q,\Lambda) - 
   Q^2\frac{\Lambda^2-\Lambda_0^2}{\Lambda^2+Q^2}
   \left[1+x+x^2+\dots\right]
\nonumber\\ &=& 
   \frac{\Lambda^2}{\Lambda^2+Q^2} -
   Q^2\frac{\Lambda^2-\Lambda_0^2}{\Lambda^2+Q^2}
   \frac1{1-x}
\nonumber\\ &=& 
   \frac{\Lambda^2}{\Lambda^2+Q^2} +
   \frac{\Lambda^2_0}{\Lambda^2_0+Q^2} -
   \frac{\Lambda^2}{\Lambda^2+Q^2}  
\nonumber\\ &=& 
   \frac{\Lambda^2_0}{\Lambda^2_0+Q^2} 
\;,\end{eqnarray} 
which is \emph{manifestly independent of $\Lambda$}.
This result represents a beautiful and pedagogic 
example for how the renormalization group works.

\subsection{Renormalization by counter terms}
The regularization by counter terms has been discussed
in Sec.~\ref{sec:3}.
I repeat it here for clarity.
Introduce a counter term Hamiltonian by
\begin{eqnarray}
    R(Q,\Lambda) \Longrightarrow 
    \overline R(Q,\Lambda) =
    R(Q,\Lambda) + C(Q,\Lambda)
\;.\end{eqnarray} 
The counter term $C(Q,\Lambda)$ is the unknown,  
subject to the constraint
\begin{eqnarray}
    \left.\phantom{\frac{d\overline R}{d\Lambda}}
    C(Q,\Lambda)\right|_{\Lambda=\Lambda_0} =0
\;.\end{eqnarray} 
The renomalization condition, 
\begin{eqnarray}
    \frac{d\overline R(Q,\Lambda)} {d\Lambda} =0
\;,\end{eqnarray} 
defines the derivative of $C$:
\begin{equation}
    \frac{dC(Q;\Lambda)}{d\Lambda} = - \frac{dR(Q;\Lambda)}{d\Lambda} 
\;.\end{equation} 
Including the constraint by the integral form,
\begin{equation}
    C(Q,\Lambda) = - \int\limits_{\Lambda_0}^{\Lambda}
    \!\!ds\ \frac{dR(Q,s)}{ds} = 
    R(Q,\Lambda_0) - R(Q,\Lambda) 
\;,\end{equation}
determines the counter term.
The renormalized regulator 
\begin{equation}
    \overline R(Q,\Lambda) = R(Q,\Lambda_0)
\end{equation}
is then \emph{manifestly independent of} $\Lambda$,
in analogy to Eq.(\ref{eq:10}).

\subsection{Renormalization by the subtraction method}
After my talk at one of the Trento meetings, 
Tobias Frederico approached me 
with the proposal to put $R(q)=1$ in Eq.(\ref{eq:29}),
\textit{i.e.}
\begin{eqnarray}
    U(\vec q) &=&  -\frac{\alpha}{2\pi^2} 
   \left(\frac{4m^2}{q^2}+2\right)
   \frac{1}{4m^2}
\;,\end{eqnarray}
take the Fourier transform according to Eq.(\ref{eq:30}),
and define this way precisely the ill-defined Hamiltonian 
of the `Coulomb plus delta' interaction according to Eq.(\ref{eq:31}),
\textit{i.e.}
\begin{eqnarray}
   H &=&  \frac{\vec p ^2}{2m_r} -\frac{\alpha}{r} +
   \mu\delta^{(3)}(\vec r)
\;.\end{eqnarray}
Such a problem could then be solved by his `subtraction method,'
see \textit{f.e.} Frederico \textit{et al} \cite{FrederFrePau02},
a method based on a T-matrix approach like in scattering theory.

Frederico's initiative has put us into 
``the unique position to compare two drastically
different schemes, both conceptually and numerically, 
and show that they agree'' (citation from \cite{FrederFrePau02}).

This is of course no proof, but:
`A reasonable man is satisfied with evidence, only a fool needs proof,'
as the famous mathematician Marc Kac has said in the context 
of trying to prove Boltzman's H-theorem.

The step to the `oscillator plus delta' interaction \cite{FrePauZho02},
\begin{eqnarray}
   H &=&  \frac{\vec p ^2}{2m_r} -a + \frac{f}{2}\,r^2 +
   \mu\delta^{(3)}(\vec r)
\;,\end{eqnarray}
was then straightforward and gave beautiful agreement with all of the
the experimental \emph{meson spectra}.
\section{Perturbative versus non-perturbative renormalization}  
\label{sec:7}
Perturbative renormalization addresses to render quantities of physical 
interest \emph{independent of the regularization scale}.
Conceptually, the scale can then be driven to infinity in the remainder
of the theory, $\Lambda\Longrightarrow\infty$,
such that it drops out from the theory.
Perturbative renormalization is adequate for scatterings problems.

Non-perturbative renormalization says: `No, don't do that. 
Do not drive $\Lambda$ to infinity.'
Set the universal renormalization scale $\Lambda_0$ in accord with experiment 
and give it the same rank as the other physical parameters of the theory,
the coupling constant $\alpha$ and the mass $m$.
Non-perturbative renormalization is adequate for bound state problems.

\section{Summary and Conclusions}
\label{sec:8}
This work is an important mile stone on the long way 
from the canonical Lagrangian for quantum chromo dynamics 
down to the composition of physical
hadrons in terms of their constituting quarks and gluons,
by the eigenfunctions of a Hamiltonian.

As part of a on-going effort, 
a denumerable number of simplifying assumptions 
had to be phrased for getting a manageable formalism \cite{Pau99b}.
Among them is the formulation of an effective interaction
by the method of iterated resolvents \cite{Pau98}.
The strongest assumption in the present work, 
the simplifying Singlet-Triplet model in Sec.~\ref{sec:4},
is also the easiest to relax, as shown in the next paper.  

The biggest progress of the present work can be found in 
Sects.~\ref{sec:2} and \ref{sec:3}.
It is related to a consistent regularization and 
renormalization of a gauge theory.
The ultraviolet divergences in gauge theory are caused less by 
the possibly large momenta of the constituent particles, 
but by the large momentum \emph{transfers} in the interaction.
In a Hamiltonian approach, such as the present, one has not
much choice other than to chop them off 
by a regulating form factor in the elementary vertex interaction. 

The form factor makes its way into a regulator function 
which suppresses the large momentum transfers 
in the Fourier transform of the Coulomb interaction.
The arbitrariness in chopping off the \emph{large momentum transfers} 
is reflected in the arbitrariness of the potential 
at \emph{small short distances}.
The arbitrariness allows for a pocket in the 
potential which binds the quarks in a hadron \cite{Pau03b}.

The problem is then how to fix this function with its
many parameters, by experiment.
In practice this is less difficult than anticipated.
Actually, it has ben done already \cite{Pau03b}.
It suffices to determine only three parameters, two
continuous ones and one counting index.

The present work opens a broad avenue of applications,
among them also the baryons and physical nuclei.
But much work must be done in the future before such a simple 
approach be taken serious.

\end{document}